\documentclass[fleqn,usenatbib]{mnras}

\usepackage{newtxtext,newtxmath}

\usepackage[T1]{fontenc}

\DeclareRobustCommand{\VAN}[3]{#2}
\let\VANthebibliography\thebibliography
\def\thebibliography{\DeclareRobustCommand{\VAN}[3]{##3}\VANthebibliography}

\usepackage{graphicx}	
\usepackage{amsmath}	

\title[Mapping the Galactic magnetic field in 3D]{Mapping the Galactic Magnetic Field Orientation and Strength in Three Dimensions}

\author[Hu \& Lazarian]{
Yue Hu$^{1,2}$\thanks{E-mail: yue.hu@wisc.edu}
, A. Lazarian$^{2,3}$\thanks{E-mail: alazarian@facstaff.wisc.edu}
\\
$^{1}$Department of Physics, University of Wisconsin-Madison, Madison, WI, 53706, USA\\
$^{2}$Department of Astronomy, University of Wisconsin-Madison, Madison, WI, 53706, USA\\
$^{3}$Centro de Investigación en Astronomía, Universidad Bernardo O’Higgins, Santiago, General Gana 1760, 8370993,
Chile\\
}
\date{Accepted XXX. Received YYY; in original form ZZZ}

\pubyear{2022}

\begin{document}
\label{firstpage}
\pagerange{\pageref{firstpage}--\pageref{lastpage}}
\maketitle

\begin{abstract}
The mapping of the Galactic Magnetic Field (GMF) in three dimensions is essential to comprehend various astrophysical processes that occur within the Milky Way. This study endeavors to map the GMF by utilizing the latest MM2 technique, the Velocity Gradient Technique (VGT), the Column Density Variance Approach, and the GALFA-HI survey of Neutral Hydrogen (HI) emission. The MM2 and VGT methods rely on an advanced understanding of magnetohydrodynamics turbulence to determine the plane-of-the-sky (POS) magnetic field strength and orientation respectively. The HI emission data, combined with the Galactic rotational curve, gives us the distribution of HI gas throughout the Milky Way. By combining these two techniques, we map the GMF orientation and strength, as well as the Alfv\'en Mach number $M_{\rm A}$ in 3D for a low galactic latitude ($b<30^{\rm o}$) region close to the Perseus Arm. The analysis of column density variance gives the sonic Mach number $M_{\rm s}$ distribution, The results of this study reveal the sub-Alfv\'enic and subsonic (or trans-sonic) nature of the HI gas. The variation of mean $M_{\rm A}$ along the line-of-sight approximately ranges from 0.6 to 0.9, while that of mean $M_{\rm s}$ is from 0.2 to 1.5. The mean magnetic field strength varies from 0.5~$\mu$G to 2.5~$\mu$G exhibiting a decreasing trend towards the Galaxy's outskirt. This work provides a new avenue for mapping the GMF, especially the magnetic field strength, in 3D. We discuss potential synergetic applications with other approaches. 
\end{abstract}




\section{Introduction}
The magnetic field is one of the most important factors in shaping the structure and evolution of the interstellar medium (ISM, \citealt{1986ApJ...301..339T,Crutcher12,2013pss5.book..641B,2017ARA&A..55..111H}). It is the source of the largest reservoir of energy in the interstellar environment, influencing the distribution and motion of interstellar gas and dust \citep{Crutcher12,BG15}, and playing a crucial role in the formation and evolution of stars and galaxies \citep{MK04,MO07,Crutcher12,2014SSRv..181....1L,2022ApJ...941...92H}. The magnetic field can also interact with cosmic rays \citep{2002PhRvL..89B1102Y,2013ApJ...779..140X,2022FrASS...9.0900B,HLX21a}, producing important ionization effects in the ISM \citep{1969ApJ...155L.149F,1981PThPS..70...35H}, and providing a means of regulating the thermal balance and overall structure of the interstellar environment \citep{2006ApJ...645L..25L,2022MNRAS.515.5267B}. 

Despite its significance, however, the magnetic field remains one of the least well-understood aspects of the ISM. One of the major difficulties is that current observational measurements, like dust polarization \citep{BG15,2015A&A...576A.104P} and Zeenman splitting \citep{Crutcher04,Crutcher12}, cannot directly trace the magnetic field in 3D space. Especially, none of them can map the magnetic field strength on the plane-of-the-sky (POS). The lack of information on POS magnetic field strength in 3D obscures our understanding of the Galactic magnetic field, but underscores the need for new and innovative observational approaches.

Based on recent advancements in the understanding of the anisotropic nature of MHD turbulence \citep{GS95,LV99}, several approaches have been proposed to map the magnetic fields \citep{2011ApJ...740..117E,2015ApJ...814...77E,GL17,2021ApJ...910...88X, 2021ApJ...915...67H}, or even trace 3D magnetic fields \citep{2021ApJ...915...67H,2023MNRAS.519.3736H}. Among these, the Velocity Gradient Technique (VGT; \citealt{GL17,LY18a,HYL18}), is the most widely used and tested \citep{HYL20,Hu19a,2020MNRAS.496.2868L,2022A&A...658A..90A,2022MNRAS.510.4952L}. Compared to dust polarization, the VGT advantageously utilizes spectroscopic observations to trace the magnetic field orientation and magnetization $M_{\rm A}^{-1}$, where $M_{\rm A}$ is the Alfv\'enic Mach number \citep{Lazarian18,HLS21}. The spectroscopic observations provide velocity information in thin velocity channel maps due to the dominated velocity caustics effect \citep{LP00}, as demonstrated in \cite{2023arXiv230610005H}. VGT's application to neutral hydrogen (HI) channel maps \citep{HYL20,2020MNRAS.496.2868L}, thus, holds the promise to map the Galactic magnetic field in 3D. With the assistance of the Galactic rotation curve, the pilot study in \cite{2019ApJ...874...25G} determined the spatial positions of HI gas and used VGT to successfully recover the magnetic field orientation traced by starlight polarization in 3D. 

However, to map a magnetic field vector, magnetic field orientation and strength are both indispensable. For the purpose of mapping the strength, \cite{2020arXiv200207996L} proposed a new technique, which is based on the use of two Mach numbers, the sonic Mach number $M_{\rm s}$ and $M_{\rm A}$. The technique is termed "MM2". The crucial information of $M_{\rm A}$ and $M_{\rm s}$ can be obtained from VGT \citep{Lazarian18} and the analysis of column density variance \citep{2012ApJ...755L..19B}, respectively. The synergetic application of these techniques to HI spectroscopic observation, therefore, can uniquely map the POS magnetic field orientation and strength in 3D. Such a magnetic field distribution in 3D is crucial for answering a diverse of astrophysical questions, including the magnetic field's role in the Galactic ecosystem, the acceleration and propagation of cosmic rays \citep{2013ApJ...779..140X,2019JCAP...05..004F,2022FrASS...9.0900B,2022FrP....10.2799L}, star formation \citep{MK04,MO07,Crutcher12,2014SSRv..181....1L}, and characterization of the CMB foreground \citep{2016A&A...594A..25P}. For this purpose, we target a low galactic latitude ($b<30^{\rm o}$) region close to the Perseus Arm and use the high-resolution Galactic Arecibo L-Band Feed Array HI (GALFA-HI) survey observed with the Arecibo 305m radio antenna \citep{2018ApJS..234....2P}. By using the Galactic rotational curve \citep{1985ApJ...295..422C}, we will show the first mapping of POS magnetic field strength, $M_{\rm A}$, and $M_{\rm s}$ in 3D.  

This paper is organized as follows. In \S~\ref{sec:data}, we briefly describe the observational data. In \S~\ref{sec:method}, we introduce the statistic tools, including VGT, column density variance approach, and MM2 used in this work. In \S~\ref{sec:result}, we present our results of magnetic field orientation, strength, $M_{\rm A}$, and $M_{\rm s}$'s distributions in 3D space. We discuss the potential uncertainties in our analysis in \S~\ref{sec:dis} and give a summary in \S~\ref{sec:con}.

\section{Observational data}
\label{sec:data}
The GALFA-HI survey is used in this work \citep{2018ApJS..234....2P}. The HI data of the Data Release 2 has a beam resolution of $4'\times4'$ (gridded into $1'\times1'$ per pixel), a spectral resolution of 0.18 km s$^{-1}$, and a brightness temperature noise of $\simeq$ 40~mK per 1 km s$^{-1}$ integrated channel.

For our analysis, we take a low galactic latitude ($b<30^{\rm o}$) region close to the Perseus Arm. The selected region stretches from Right Ascension (R.A.) $\simeq$ 73.5$^\circ$ to 63$^\circ$ and Declination (DEC.) $\simeq$ 19$^\circ$ to 24.5$^\circ$. The analysis is performed for the full velocity range -60~km/s to 30~km/s.

\begin{figure}
\centering
\includegraphics[width=1\linewidth]{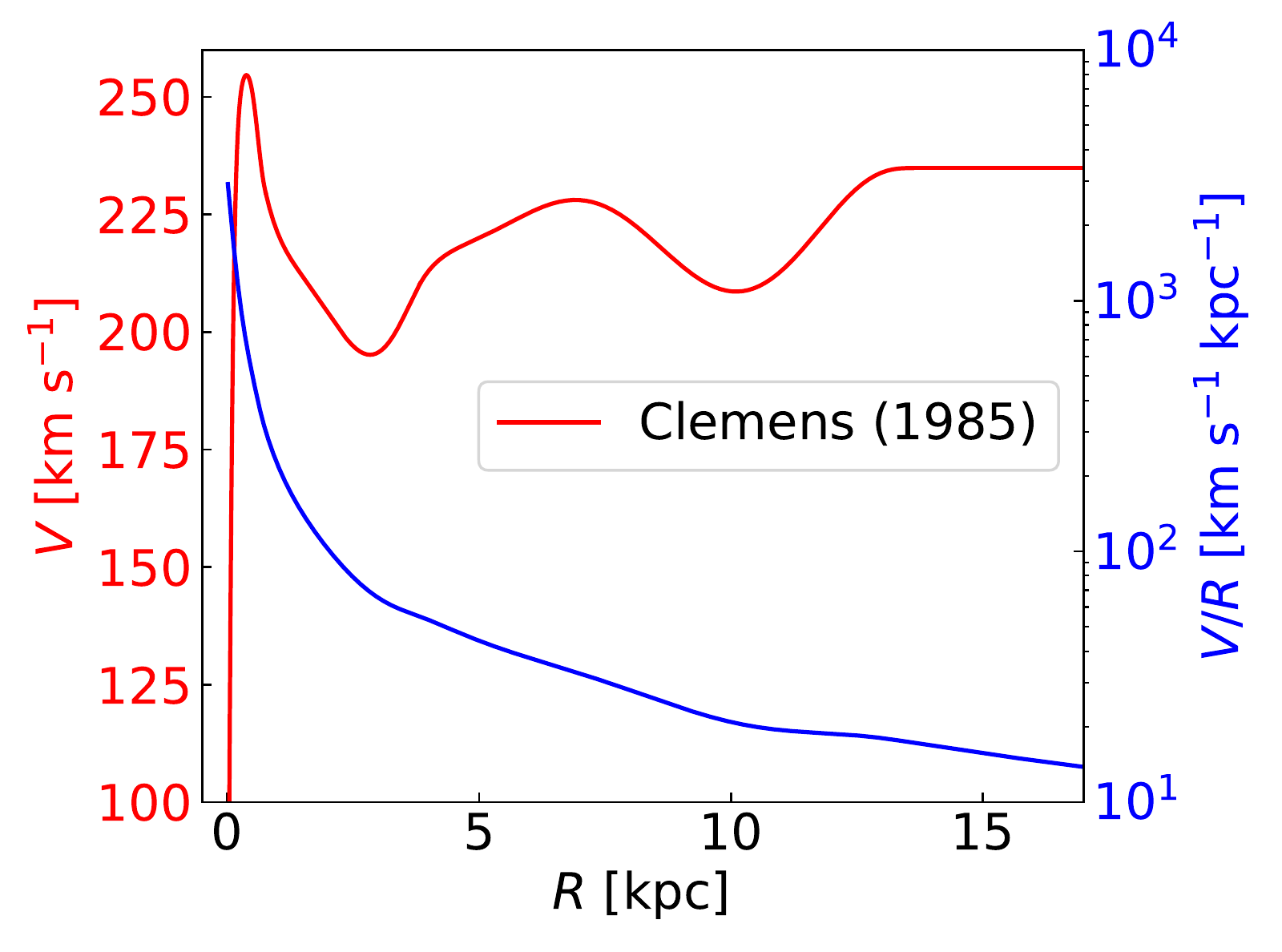}
    \caption{\textbf{Red}: Galactic rotational curve used in this work. Adopted from \citet{1985ApJ...295..422C}. \textbf{Blue}: the radio $V/R$ calculated from the Galactic rotational curve.}
    \label{fig:GCR}
\end{figure}

\begin{figure*}
\centering
\includegraphics[width=1\linewidth]{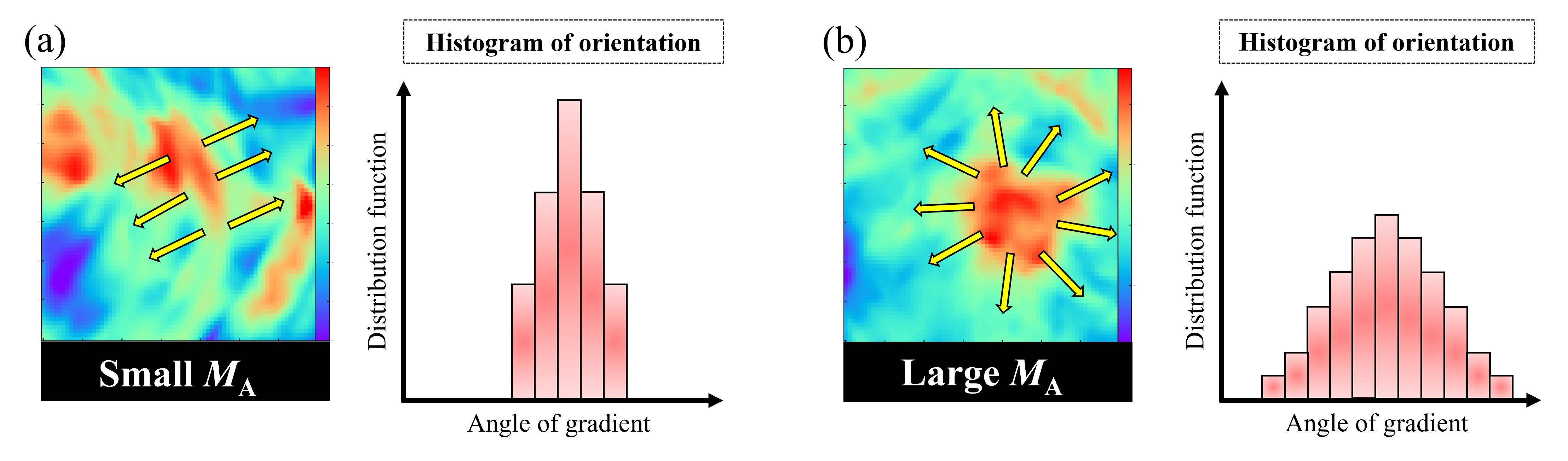}
    \caption{An illustration of how the magnetization $M_{\rm A}$ affects the gradient orientation histogram. \textbf{(a)}: the turbulent eddies in a strongly magnetized medium are highly anisotropic elongating along the magnetic fields. Their gradients (yellow arrows) orient in similar directions so that the histogram is less dispersed. \textbf{(b)}: in a weakly magnetized medium, the eddies, as well as the gradients, are more isotropic. The histogram in this case is more dispersed. In an extreme case of purely hydrodynamic turbulence, the histogram would appear a uniformly random distribution.}
    \label{fig:illustration_ma}
\end{figure*}

\begin{figure*}
\centering
\includegraphics[width=1\linewidth]{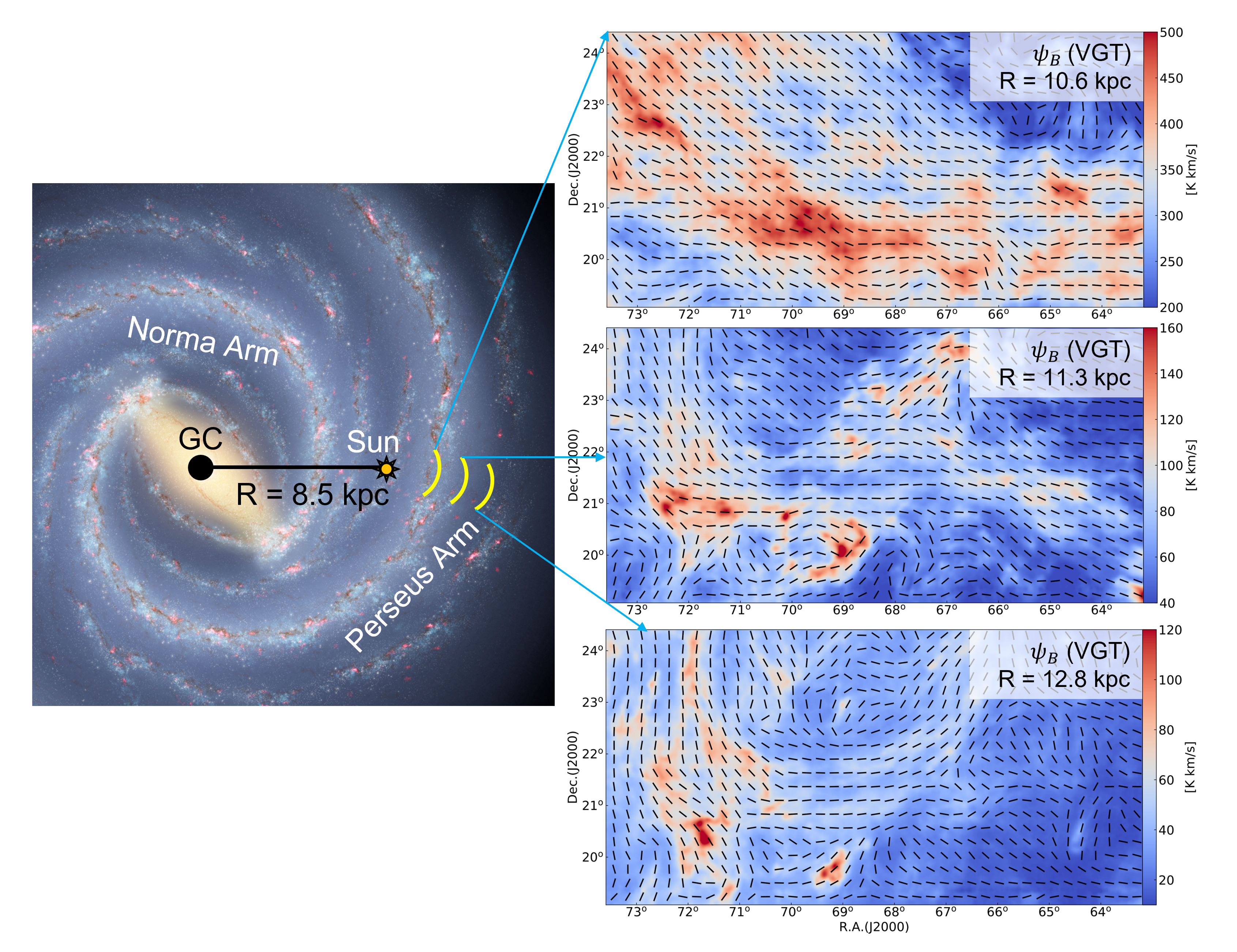}
    \caption{Maps of magnetic field orientation $\psi_B$ at three different Galactic spatial places. The magnetic field orientation (black segment) is superimposed on the integrated HI intensity maps (right) with a channel width of $\sim10$~km/s. }
    \label{fig:Bdirection}
\end{figure*}


\section{Methodology}
\label{sec:method}
\subsection{Galactic rotational curve}
An accurate rotation curve for the Milky Way Galaxy is
crucial to constrain the spatial distribution of HI gas. Here we adopt the high-order polynomial curve obtained by \citet{1985ApJ...295..422C}. The composite curve is in the form of:
\begin{equation}
\label{eq.grc}
    \begin{aligned}
    V(R)&=\sum_{i=0}^6A_iR^i,~~~R<0.09R_0,\\
    &=\sum_{i=0}^5B_iR^i,~~~0.09R_0<R<0.45R_0,\\
    &=\sum_{i=0}^7C_iR^i,~~~0.45R_0<R<1.6R_0,\\
    &=D_0,~~~1.6R_0<R,\\
    \end{aligned}
\end{equation}
where $R$ is the distance from the Galactic center to the point of interest within the Galaxy, $R_0=8.5$~kpc is the distance from the center of our Galaxy to the Sun, and $V$ is the circular velocity of the point. $A,B,C,D$ are the fitted coefficients of the curve assuming the Sun's circular velocity $V_0=220$~km/s. The coefficients are given in Tab.3 in \citet{1985ApJ...295..422C}. Instead of listing the numerous coefficients here, we reproduced the plot of the rotational curve in Fig.~\ref{fig:GCR}. After $1.6R_0$, we assume a flat curve. 

With the curve, we can determine the spatial distribution of HI gas using the relative velocity $V_r$ and angular velocity $\omega=V/R$ (see Fig.~\ref{fig:GCR}) of Galactic rotation at the point. Their relation is expressed as \citep{2007ApJ...671..427M}:
\begin{equation}
    V_r=R_0(\frac{V}{R}-\frac{V_0}{R_0})\sin l \cos b,
\end{equation}
where $l$ is the Galactic longitude and $b$ is the latitude. For the small patch of sky considered in this work, we adopt $l$ and $b$ from the region's central coordinates. $V_r$ is determined by the central velocity of every thick HI channel (i.e., channel width $\sim10$~km/s). The reason for choosing a $\sim10$~km/s channel width is given in \S~\ref{subsec:vgt}.

\subsection{Velocity gradient technique}
\label{subsec:vgt}
\subsubsection{Mapping magnetic field orientation}
VGT is developed from MHD turbulence's anisotropy to map the magnetic field in ISM. The anisotropy means the magnetized and turbulent eddies are anisotropic, i.e., elongating along the magnetic field \citep{GS95,LV99}. Consequently, the maximum velocity fluctuation, as well as the gradient of velocity fluctuation, appears (or orients) in the direction perpendicular to the magnetic field (see \citealt{LY18a} and \citealt{HLY20} for details).

Obtaining information on velocity fluctuations from observations is difficult. One of the possible ways is using the effect of velocity caustics \citep{LP00,2017MNRAS.464.3617K}. The velocity caustics effect means, instead of revealing the real density structures in space, the observed intensity structures in a velocity channel are distorted due to turbulence which raises different velocities along the LOS. The significance of velocity caustics is correlated with the channel width of spectroscopic observation. When a velocity channel is extremely thick, the integrated intensity map along the LOS essentially recovers column density. When the velocity channel is narrow enough (i.e., the channel width is smaller than the velocity dispersion), the effect and distortion become significant. The observed intensity fluctuation, in this case, is dominated by velocity fluctuation rather than density fluctuation \citep{LP00}. One can, therefore, use the thin velocity channels to get velocity fluctuations and calculate the velocity gradient.

\cite{2019ApJ...874..171C} questioned the validity of velocity caustics with the presence of thermal broadening in multi-phase HI gas and proposed that the thin velocity channel is dominated by density fluctuation. The questions were answered in \cite{2019arXiv190403173Y}. Especially, \cite{2021ApJ...910..161Y} proposed the Velocity Decomposition Algorithm (VDA) to separate velocity and density contributions in a velocity channel. The application of VDA to the GALFA-HI survey and the multi-phase HI simulations demonstrated the dominance of velocity fluctuation in a thin velocity channel \citep{2023arXiv230610005H}. 

In view of that the typical value of turbulent velocity dispersion in a 100~pc cloud is $\sim10$~km/s \citep{2022ApJ...934....7H}, we adopt the value to define the width of a thick HI channel (corresponding to an HI cloud), i.e., $\sim10$~km/s. The GALFA-HI data's narrowest channel width of $\sim0.184$~km/s is used to define a thin channel. 

For the VGT calculation, we adopt the recipe used in \cite{2022MNRAS.511..829H} to calculate the velocity gradient. The calculation is conducted by: (i) convolving each thin channel map with 3×3 Sobel Kernels to create a raw gradient map (pixels are blanked out if their intensity is less than three times the noise level); (ii) averaging the raw gradient angle map over each sub-block of $16\times16$ pixels, which guarantee statistically sufficient samples. The average is performed by fitting the gradient angle's histogram with a Gaussian distribution and then finding the most probable orientation, which is taken as the mean gradient angle for that sub-block \citep{YL17a}; (iii) constructing pseudo-Stokes parameters $Q_{\rm g}$ and $U_{\rm g}$. The cosine and sine values of each sub-block averaged gradient map is weighted with the corresponding channel's intensity and integrated along the LOS to get $Q_{\rm g}$ and $U_{\rm g}$, respectively. (iv) getting the POS magnetic field orientation from $\psi_B=\frac{1}{2}\arctan(U_{\rm g},Q_{\rm g})+\frac{\pi}{2}$. Note here the gradient is calculated for every thin channel. The pseudo-Stokes parameters, however, are constructed for every thick channel, which means the velocity range for integrating the (thin channel) gradient is $\sim10$~km/s. In combining with the Galactic rotational curve, we can locate the spatial position of every thick channel and get the maps of magnetic field orientation in 3D.

\subsubsection{Mapping Alfv\'en Mach number}
In addition, VGT is also capable to derive the magnetization, i.e., $M_{\rm A}^{-1}$, where $M_{\rm A}$ is the Alfv\'en Mach number \citep{Lazarian18}. This approach is also based on the anisotropy of MHD turbulence. As illustrated in Fig.~\ref{fig:illustration_ma}, in a strongly magnetized medium (i.e., small $M_{\rm A}$), the turbulent eddies are highly anisotropic being elongated along the magnetic field. The anisotropy, however, is less significant in a weakly magnetized medium. As the gradient is perpendicular to the eddy's structure, this change of anisotropy can be detected by the gradient orientation's histogram. The histogram is less dispersed in the case of small $M_{\rm A}$, while the distribution spreads wilder when $M_{\rm A}$ increases. The dispersion of the histogram is characterized by the so-called “Top-to-Bottom” ratio of the distribution. The correlation with $M_{\rm A}$ is given as \citep{Lazarian18}:
\begin{equation}
\label{eq.tb}
    \begin{aligned}
       M_{\rm A}&\approx1.6 (T/B)^{\frac{1}{-0.60\pm0.13}}, M_{\rm A}\le1,\\
    M_{\rm A}&\approx7.0 (T/B)^{\frac{1}{-0.21\pm0.02}},  M_{\rm A}>1,\\
    \end{aligned}
\end{equation}
where $T$ denotes the maximum value of the fitted histogram of the velocity gradient's orientation, while $B$ is the minimum value. Based on Eq.~\ref{eq.tb}, we take raw velocity gradients with every sub-block of $16\times16$ pixels to calculate the $T/B$ ratio and derive $M_{\rm A}$ accordingly.

\begin{figure*}
\centering
\includegraphics[width=1\linewidth]{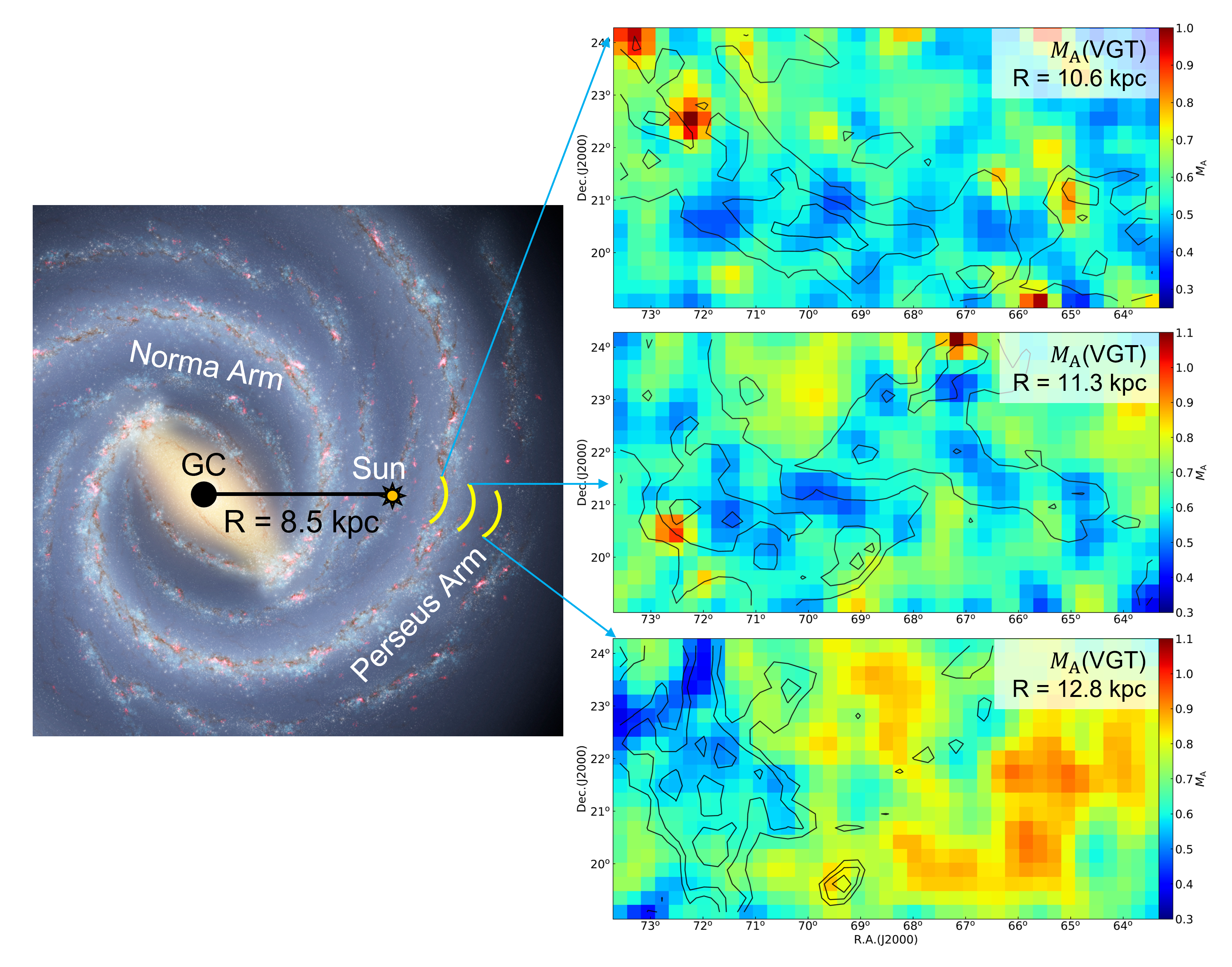}
    \caption{Same as Fig.~\ref{fig:Bdirection}, but for the magnetization $M_{\rm A}$. The contours outline the prominent HI intensity structures (see Fig.~\ref{fig:Bdirection}) in each map.}
    \label{fig:Ma}
\end{figure*}

\begin{figure*}
\centering
\includegraphics[width=1\linewidth]{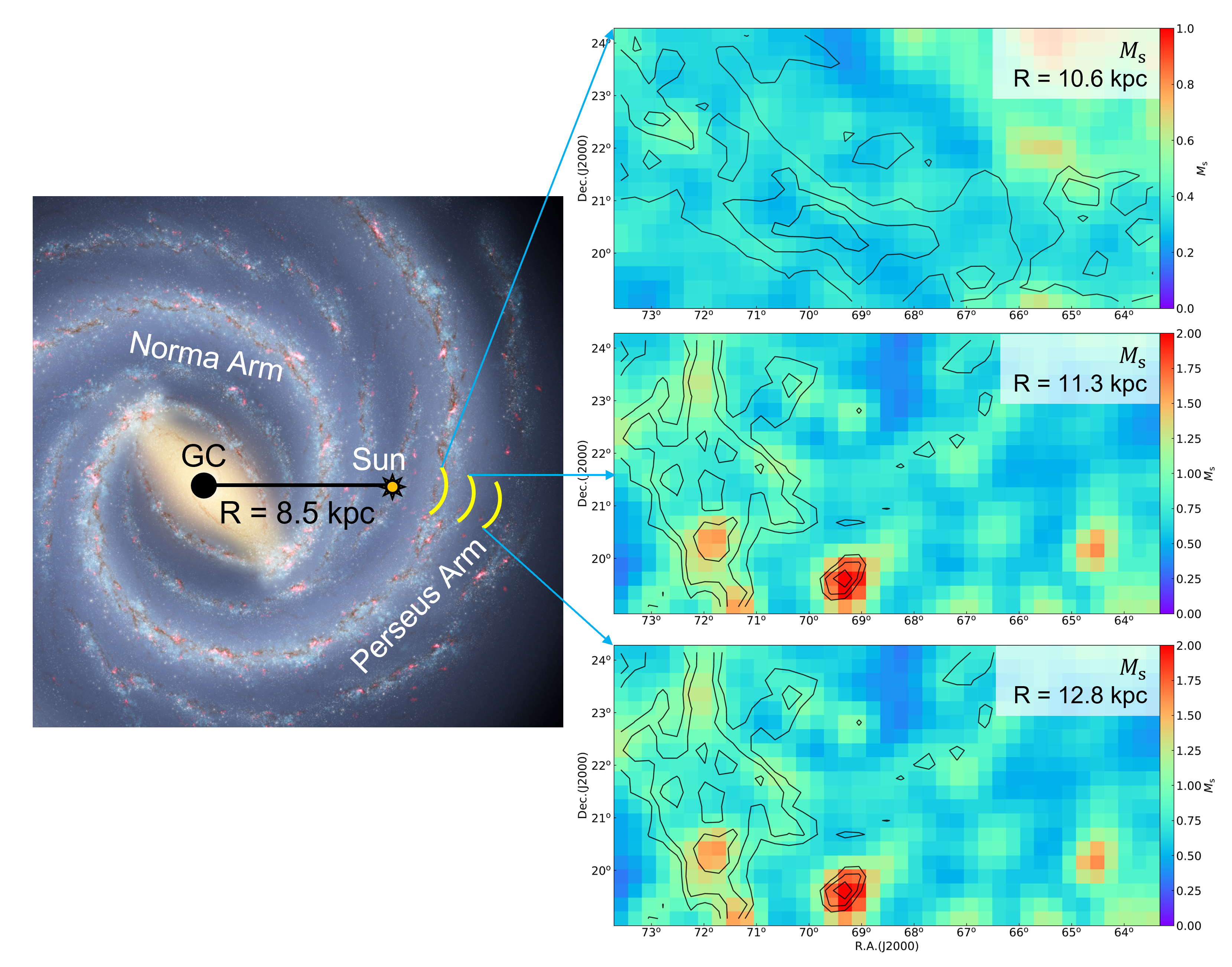}
    \caption{Same as Fig.~\ref{fig:Bdirection}, but for the sonic Mach number $M_{\rm s}$. The contours outline the prominent HI intensity structures (see Fig.~\ref{fig:Bdirection}) in each map.}
    \label{fig:Ms}
\end{figure*}

\begin{figure*}
\centering
\includegraphics[width=1\linewidth]{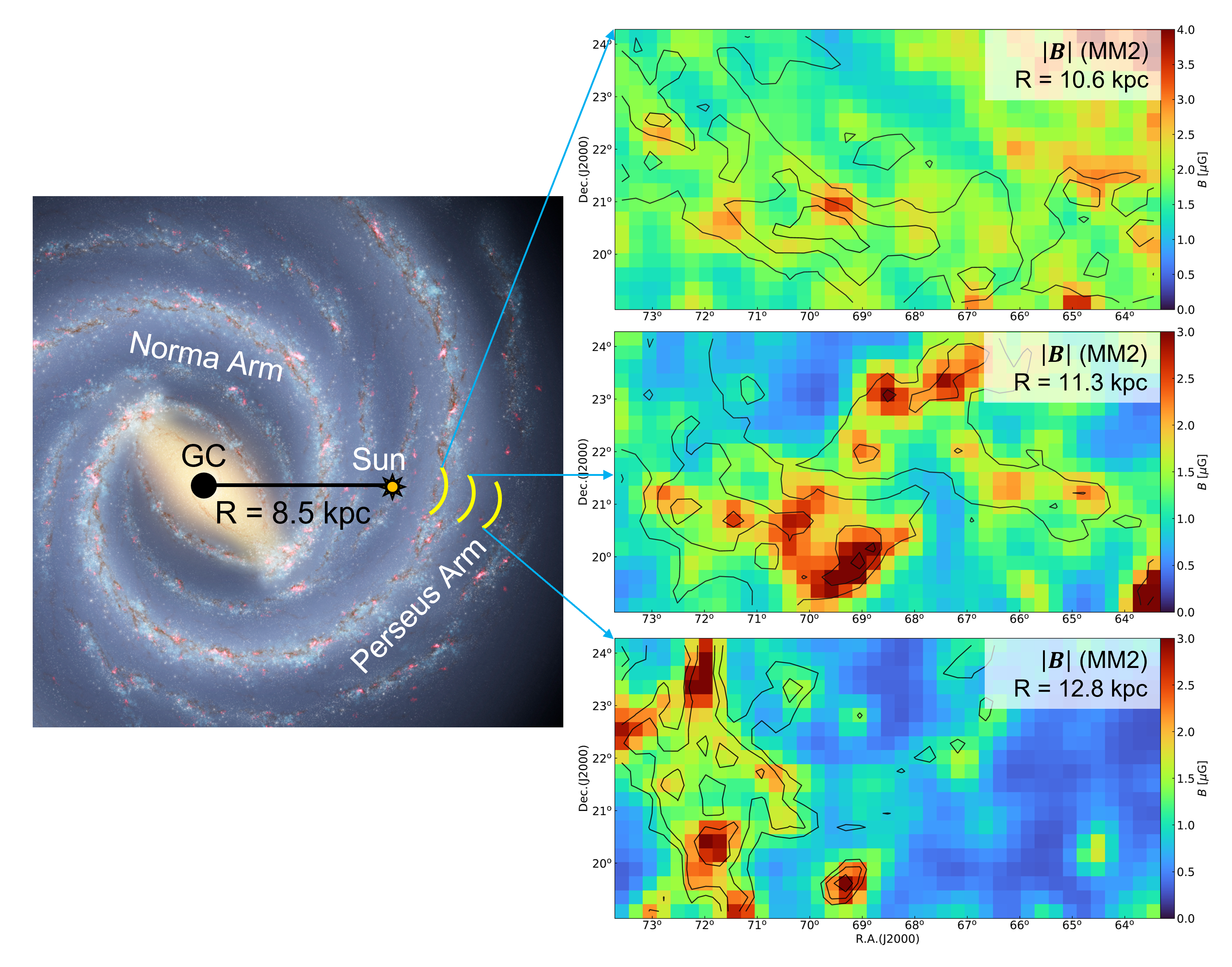}
    \caption{Same as Fig.~\ref{fig:Bdirection}, but for POS magnetic field strength $|\pmb{B}|$. The contours outline the prominent HI intensity structures (see Fig.~\ref{fig:Bdirection}) in each map.}
    \label{fig:Bstrength}
\end{figure*}

\begin{figure}
\centering
\includegraphics[width=1\linewidth]{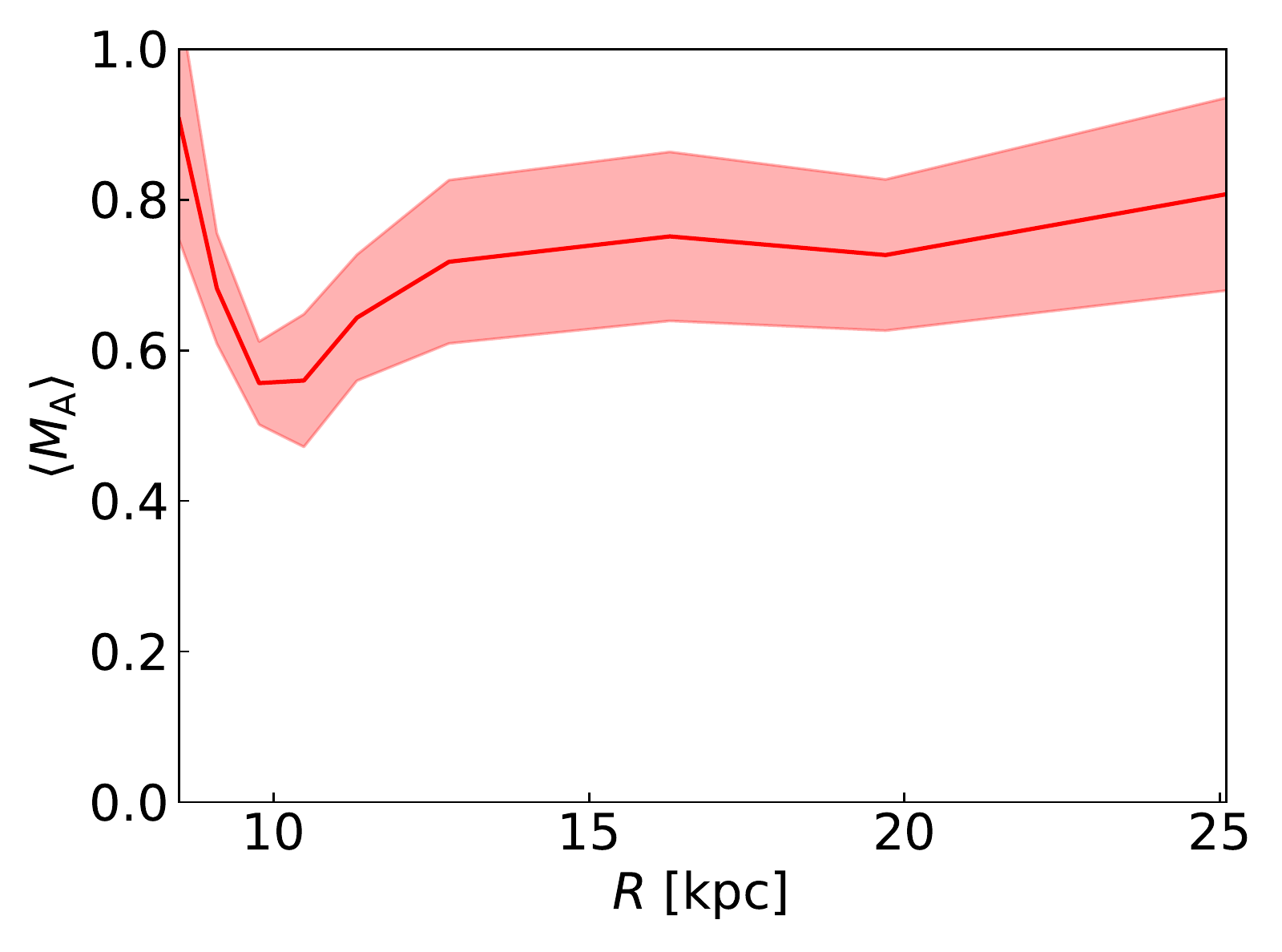}
    \caption{The mean Alfv\'en number $\langle M_{\rm A}\rangle$ as a function of distance $R$ to the Galactic center. The red shadow area indicates the range of uncertainty given by the standard deviation.}
    \label{fig:MaR}
\end{figure}

\begin{figure}
\centering
\includegraphics[width=1\linewidth]{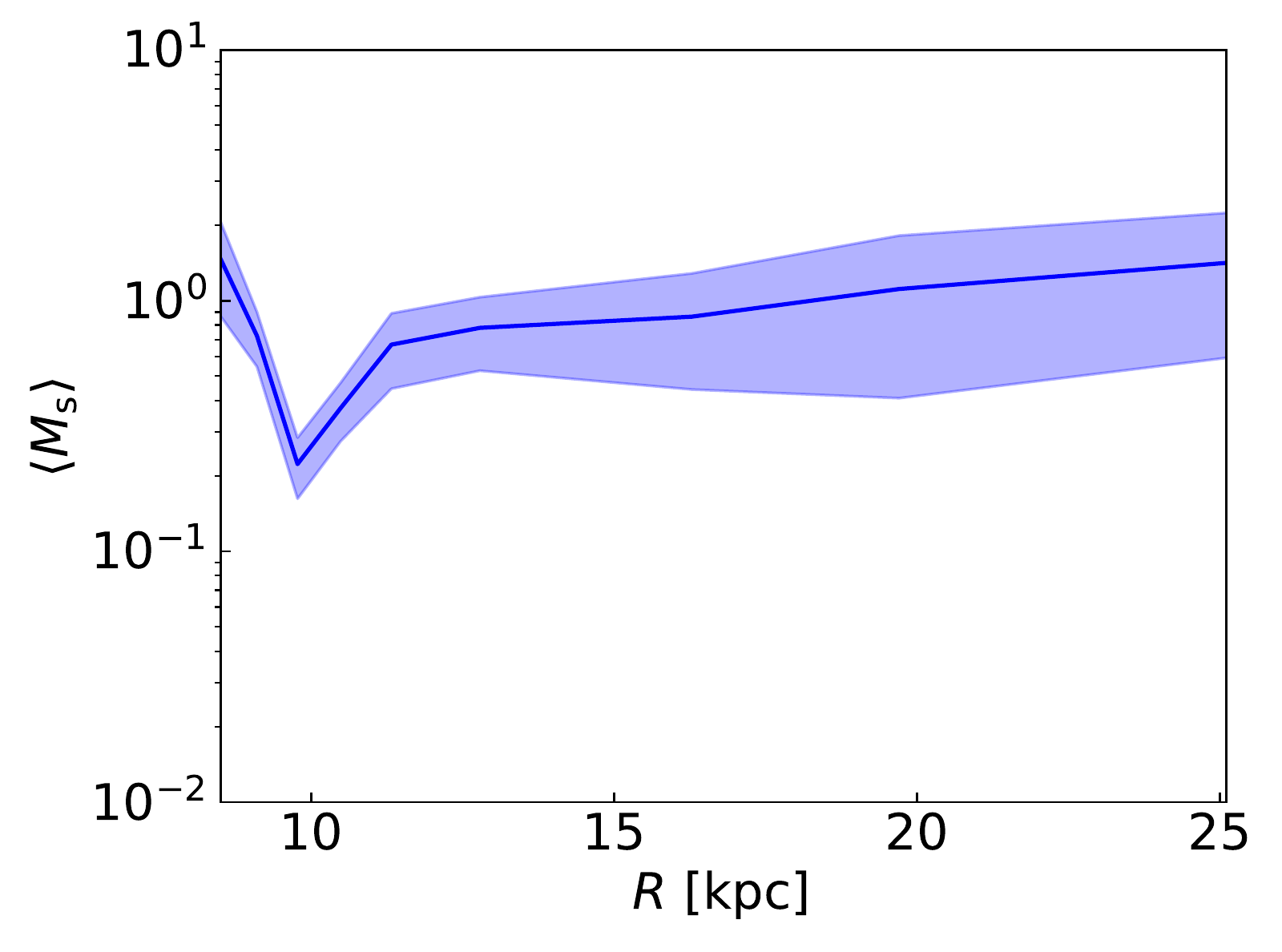}
    \caption{The mean sonic number $\langle M_{\rm s}\rangle$ as a function of distance $R$ to the Galactic center. The blue shadow area indicates the range of uncertainty given by the standard deviation.}
    \label{fig:MsR}
\end{figure}

\begin{figure}
\centering
\includegraphics[width=1\linewidth]{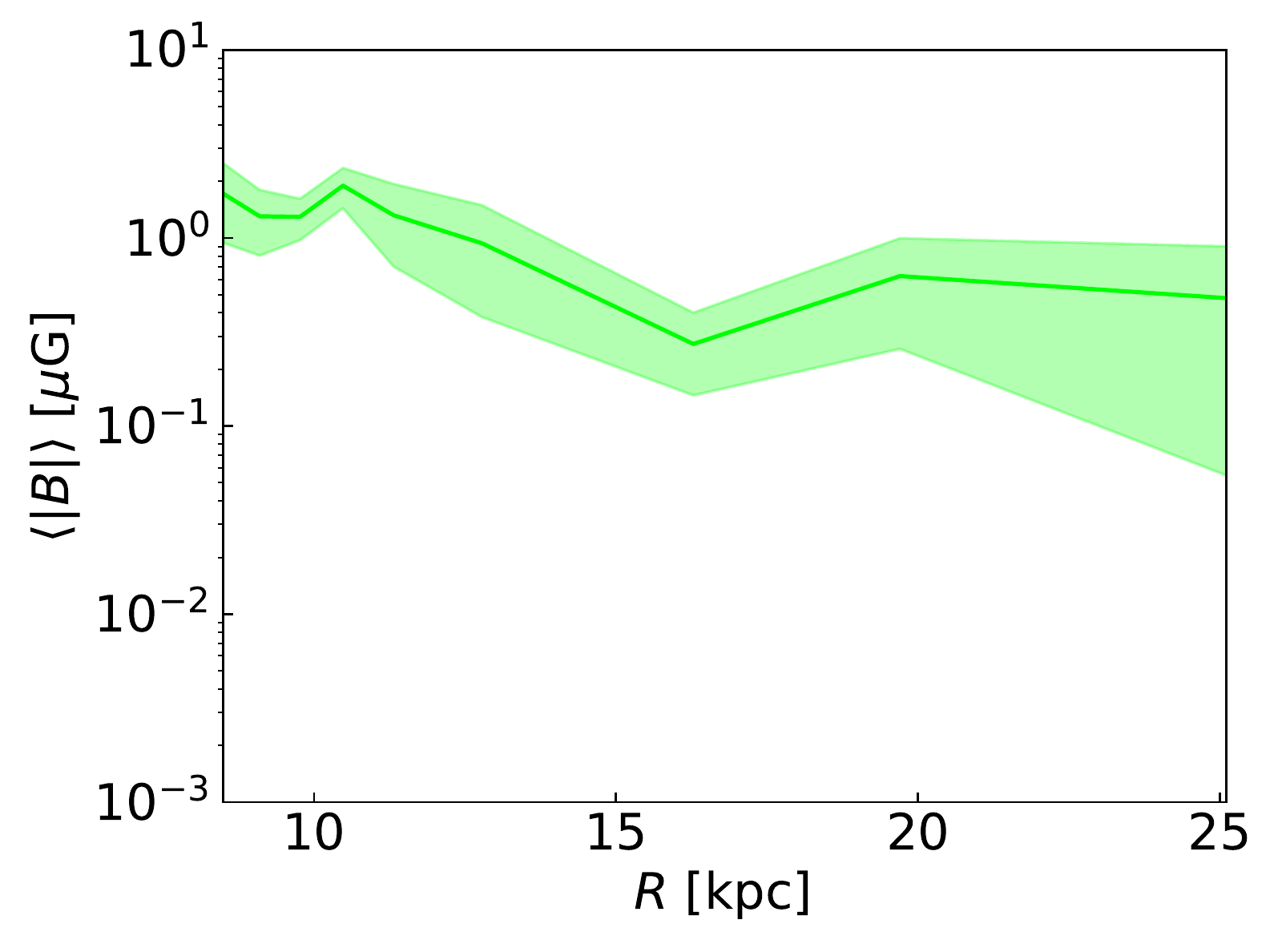}
    \caption{The mean POS magnetic field strength $\langle |\pmb{B}|\rangle$ as a function of distance $R$ to the Galactic center. The lime shadow area indicates the range of uncertainty given by the standard deviation.}
    \label{fig:BR}
\end{figure}
\subsection{Column density variance \& $M_{\rm s}$}
The sonic Mach number is defined as the ratio of turbulent velocity dispersion $\sigma_v$ and sound speed $c_s$, i.e., $M_{\rm s}=\sigma_v/c_s$. Typically $\sigma_v$ can be derived from linewidth for a molecular cloud. This is, however, more difficult for HI gas, due to  a significant line broadening arising from regular motions, e.g. differential rotation, which is not related to turbulence.

Nevertheless, \cite{2012ApJ...755L..19B} proposed a new way to map $M_{\rm s}$ based on column density statistics. This method is based on the fact that density fluctuation's amplitude increases for a large $M_{\rm s}$. This increase in fluctuation results in a higher variance. Accordingly, \cite{2012ApJ...755L..19B} showed the relation of $M_{\rm s}$ and the variance $\sigma^2_{N/N_0}$ of the normalized column density field $N/N_0$ (where $N_0$ is the mean column density) is:
\begin{equation}
    \sigma^2_{N/N_0}=(\frac{M_{\rm s}^2}{9}+1)^{0.11}-1.
\end{equation}
To find the HI column density of every thick HI channel, we adopt the conversion from the HI brightness temperature $T_{\rm MB}$ to column density \citep{2019ApJ...872...56P}:
\begin{equation}
   N=\int_{v_0-5}^{v_0+5} 1.823\times10^{18}T_{\rm MB} dv,
\end{equation}
where $N$ denotes the column density of HI in the unit of $\rm{cm^{-2}}$, $dv$ is the spectral resolution in the unit of km/s, and $v_0$ is the thick channel's central velocity in the unit of km/s. Then the variance and $M_{\rm s}$ are calculated for every sub-block of $16\times16$ pixels, which is selected to match VGT's resolution.

\subsection{MM2 technique}
The Davis–Chandrasekhar–Fermi (DCF) method \citep{1951PhRv...81..890D,1953ApJ...118..113C} is widely used to estimate the POS magnetic field strength \citep{HLS21,2021ApJ...907...88P,2021ApJ...913...85H,2021MNRAS.tmp.3119L,2022ApJ...929...27H,2022arXiv220512084T}. It assumes magnetic field fluctuations $\delta B$ are fully produced by velocity fluctuations $\sigma_v$ of Alfv\'enic turbulence. In this case, the kinetic energy of turbulence is completely transferred to the energy of magnetic field fluctuations:
\begin{equation}
\label{eq:6}
    \frac{1}{2}\rho\sigma_v^2=\frac{1}{8\pi}\delta B^2,
\end{equation}
where $\rho$ is the gas mass density. 

For observational implementation, $\delta B$ can be approximately obtained from the fluctuations of magnetic field angle: $\delta \psi\sim\delta B/B$, where $B$ is the mean magnetic field \citep{1996ASPC...97..486Z}, so one can get:
\begin{equation}
    B\approx f\frac{\sqrt{4\pi\rho}\sigma_v}{\delta\psi},
\end{equation}
here $f$ is a correction factor and $\sigma_v$ is typically estimated from line width. However, this approach to obtaining $\delta B$ and $\sigma_v$ would overestimate $B$, when the regions under study are only small patches of the cloud. The overestimation comes from the fact that (i) $\sigma_v$ estimated from line width corresponds to the velocity fluctuation at injection scale $L_{\rm inj}$, but (ii) the $\delta \psi$ for a small patch with size $l$ corresponds to the fluctuation at scale $l$. Compared to the angle dispersion for the entire cloud (i.e., at turbulence injection scale $L_{\rm inj}$), the dispersion for a small patch is reduced by a factor of $\sim(l/L_{\rm inj})^{1/3}$, assuming Kolmogorov-type turbulence. This overestimation exists also for other modified DCF methods if $\delta \psi$ is derived at scales smaller than $L_{\rm inj}$ \citep{2021ApJ...913...85H,2021A&A...656A.118S,2022MNRAS.514.1575C}. Nevertheless, \cite{2022ApJ...935...77L} proposed the Differential Measure Approach (DMA) to address the overestimation, as observationally demonstrated in \cite{2022arXiv221011023H}.

Furthermore, in this work, we intend to use another independent approach to estimate $B$. Using $M_{\rm A}=\delta B/B$ and $M_{\rm s}=\sigma_v/c_s$, Eq.~\ref{eq:6} can be expressed as \citep{2020arXiv200207996L}:
\begin{equation}
\label{eq.mm2}
    B=c_s\sqrt{4\pi\rho}{M_{\rm s} M_{\rm A}^{-1}},
\end{equation}
which suggests that the mean POS magnetic field strength can be obtained more accurately if we know $M_{\rm A}$ and $M_{\rm s}$ at the same scales. The sound speed is $c_s=\sqrt{\frac{\gamma k_B T}{m}}$, where $k_B$ is the Boltzmann constant, $T$ is gas temperature, $\gamma$ is the adiabatic index, and $m$ is the mass of a single atom. For simplicity, we assume that the HI gas is predominantly composed of its warm and unstable phases with a temperature of $T=10000$ K. This assumption is made due to several reasons. Firstly, the volume filling factor of gas in the cold phase is found to be lower than 10\% \citep{2021arXiv211106845H}. Secondly, we derived $M_{\rm s}$ from the variance of the HI column density, and $c_s$ is used in conjunction with $M_{\rm s}$ to obtain information about velocity fluctuations. \cite{2018A&A...619A..58K} showed that within a velocity channel width of 16 km/s (similar to our case of ~10 km/s), the column density from the warm neutral medium can be one order of magnitude larger than that of the cold neutral medium. Therefore, it is natural to focus on the warm phase, which dominates the HI column density.

\section{Results}
\label{sec:result}
\subsection{Magnetic field orientation}
VGT's ability to map magnetic field orientation in diffuse ISM has been demonstrated in \cite{YL17a,HYL20}. There the VGT measurement is fully integrated with the LOS rather than separating the magnetic field in velocity space. The magnetic field orientation derived from VGT shows excellent agreement with Planck polarization at 353 GHz, which also measures the fully integrated magnetic field. \cite{2019ApJ...874...25G} further tested the possibility to decompose the VGT measurement in 3D with the assistance of the Galactic rotational curve. They showed the statistical agreement of VGT and the magnetic field inferred from star-light polarization at different Galactic spatial positions. 

In this work, as shown in Fig.~\ref{fig:Bdirection}, we present the magnetic field orientation mapped with VGT for three HI clouds (i.e., the thick channel with a width of $\sim10$~km/s) at different spatial places ($R=10.6, 11.3$, and $12.8$~kpc) in the Galaxy. The magnetic field maps are smoothed to the spatial resolution of $30'$. The intensity structures in the three clouds exhibit apparent differences and the magnetic field generally follows the structures. Other HI clouds' magnetic field maps are presented in Appendix~\ref{ap:A} (see Fig.~\ref{fig:Bdirection2}). 

\subsection{$M_{\rm A}$ and $M_{\rm s}$}
Using VGT and the density-variance approach (see \S~\ref{sec:method}), we map the distribution of $M_{\rm A}$ and $M_{\rm s}$ for the same three HI clouds, as shown in Fig.~\ref{fig:Ma} and Fig.~\ref{fig:Ms}. The maps have spatial resolution of $30'$. Other HI clouds' $M_{\rm A}$ and $M_{\rm s}$ maps are presented in Appendix~\ref{ap:B} (see Figs.~\ref{fig:Ma2} and \ref{fig:Ms2}).

We find the clouds are globally sub-Alfv\'enic with $M_{\rm A}<1$. This finding agrees with the results in \cite{2022arXiv220311179P}, which shows the ISM is generally sub-Alfv\'enic up to the hydrogen volume density of $n_{\rm H}\sim10^7~{\rm cm^{-3}}$. In addition, the clouds are globally subsonic $M_{\rm s}<1$ or trans-sonic $M_{\rm s}\sim 1$. This can be understood based on the fact that HI gas is dominated by warm or unstable phases \citep{2021arXiv211106845H}. The high temperature in the two phases results in a large thermal sound speed. In this case, $M_{\rm s}$ is not expected to be much larger than the unity. 

\subsection{POS magnetic field strength}
With the knowledge of $M_{\rm A}$ and $M_{\rm s}$ distributions, one can the distribution of POS magnetic field strength accordingly using the MM2 approach (see \S~\ref{sec:method}). The information on gas mass density is derived from the HI column density. The LOS thickness of every HI cloud is given by its spatial separation between it and another cloud located behind it. 

The three maps of POS magnetic field strength are presented in Fig.~\ref{fig:Bstrength}. Other HI clouds' $B$ maps are presented in Appendix~\ref{ap:A} (see Fig.~\ref{fig:Bstrength2}). The MM2 measurements reveal that the mean POS magnetic field strength is on the order of $\mu$G. Several strong magnetic fields $>3\mu$G regions are noticed in the regions with prominent HI intensity structures, i.e., higher density. The earlier Zeeman measurement found the LOS magnetic field strength varies from $0.1\mu$G to 10$\mu$G for diffuse ISM \citep{Crutcher12,2022arXiv220311179P}, with a median value of $6~\mu$G for cold HI \citep{2005ApJ...624..773H}. Our results statistically are about a factor of 2 lower than the Zeeman results. However, one should note that typically Zeeman measurements provide only upper limits of magnetic field strength, and dense cold HI usually is associated with relatively strong magnetic fields. Thus the actual Zeeman measurements might be biased towards places with stronger magnetic fields.

\subsection{Variation of $M_{\rm A}$, $M_{\rm s}$, and POS magnetic field strength along the LOS}
VGT, density-variance, and MM2 methods uniquely map the distribution of $M_{\rm A}$, $M_{\rm s}$, and POS magnetic field strength. With the HI clouds' spatial positions determined by the Galactic rotational curve, it is possible to investigate the variation of mean $M_{\rm A}$, $M_{\rm s}$, and magnetic field strength along the LOS. The variation is presented in Figs.~\ref{fig:MaR}, \ref{fig:MsR}, and \ref{fig:BR}. The diffuse HI gas is generally sub-Alfv\'enic and sub-sonic (or trans-sonic). $M_{\rm A}$ varies from $\sim 0.6$ to $\sim0.9$. An apparent low $M_{\rm A}$ value appears at $R\sim10$~kpc. The corresponding HI cloud exhibits very prominent HI structures with high intensity and density (see Fig.~\ref{fig:Bdirection}). One possible reason is that the high-density structures are expected to associate with a relatively strong magnetic field, so that $M_{\rm A}$ decreases. Another possibility is that velocity dispersion at this position decreases, as a drop of $M_{\rm s}$ is also observed at $R\sim10$~kpc. Unless the gas temperature varies significantly, the drop of $M_{\rm s}$ suggests a smaller velocity dispersion and a smaller $M_{\rm A}$. Nevertheless, $M_{\rm A}$ and $M_{\rm s}$ tend to be stable at $0.7$ and $0.8$, respectively, when $R>12$~kpc.

However, unlike $M_{\rm A}$ and $M_{\rm s}$, the mean magnetic field strength decreases when $R>11$~kpc. The peak value of $\sim2.5\mu$G appears at $R\sim10.5$~kpc. We expect the decrease to be mainly raised by density variation. At the outskirts of the Galaxy, HI gas is less abundant so $B\propto\sqrt{4\pi\rho}$ decreases.

\section{Alternative approaches and further work}
The advances in understanding MHD turbulence (see monograph by \citealt{2019tuma.book.....B}, review by \citealt{2013SSRv..178..163B}) open wide avenues for obtaining from observations the key parameters required for the present study. Below we present a  brief review.

\subsection{Alternative ways of obtaining $M_A$}
The anisotropy of MHD turbulence, i.e., velocity statistics, was proposed as a technique for magnetic field studies by \cite{2002ASPC..276..182L}. The anisotropy in a thin velocity channel map was demonstrated to represent the magnetic field and reflect media magnetization \citep{LY18a,Lazarian18}. Similarly, velocity centroids were studied to obtain $M_{\rm A}$ in \cite{2005ApJ...631..320E} and applied to the studies of magnetic field direction in \cite{2008ApJ...680..420H}. Further numerical studies of the anisotropy arising from the magnetic field and their relation to $M_{\rm A}$ were performed in \cite{2011ApJ...740..117E} using velocity centroids and \cite{2015ApJ...814...77E} for thin velocity channel maps. A reliable determining of magnetization for $M_{\rm A}\le 1.5$ was reported. The procedures for increasing the reliability of determining $M_{\rm A}$ in the presence of density inhomogeneities, which are correlated with $M_{\rm s}$, were explored. As a result, it was found that $M_{\rm A}$ can be obtained reliably with only marginal influence from $M_{\rm s}$.

The advances in the understanding of turbulence statistics Position-Position-Velocity (PPV) coordinates in \cite{LP00} allowed further theoretical advances in \cite{2016MNRAS.461.1227K,2017MNRAS.464.3617K} that dealt, respectively, with channel maps and velocity centroid statistics. There the observable dipole moments of structure functions of anisotropy were analytically related to the values of $M_{\rm A}$. These results were confirmed numerically by \cite{2020ApJ...901...11H}, opening the avenue for theory-based expressions to be applied to the analysis of the observational data. 

Analytical ones complemented the numerical studies in the direction above. The study of the magnetic field in diffuse media via measuring the anisotropy via structure functions, i.e.,  the Structure-Function Analysis (SFA; \citealt{HXL21,XH21,HLX21a}) provides an alternative, but synergetic, way of analyzing the data for obtaining media magnetization and the POS directions of the magnetic field. In this paper, we adopted, however, the way of obtaining magnetization that is based on velocity gradients as this approach allows more detailed maps of $M_{\rm A}$ distribution \citep{Lazarian18,Hu19a}. We plan to explore the synergy of the two approaches in our future publications.

Other tools for obtaining $M_{\rm A}$ from observational data include Tsallis statistics \citep{2010ApJ...710..125E} and bispectrum \citep{2009ApJ...693..250B}. The former was correlated with both $M_{\rm A}$ and $M_{\rm s}$, and the interplay of the two key parameters was explored in \cite{2018MNRAS.475.3324G}. In the present study, we estimate both $M_{\rm A}$ and $M_{\rm s}$. Therefore, cross-checking our findings with independent approaches opens a promising avenue for improving accuracy in determining $M_{\rm A}$ and $M_{\rm s}$, and the magnetic field strength $B$.

\subsection{Alternative ways of obtaining $M_{\rm s}$}
Our study of the $M_{\rm s}$ in this paper is based on the ways explored in \cite{2012ApJ...755L..19B}. 
However, $M_{\rm s}$ can be obtained with other approaches as well. For instance, the skewness and kurtosis of column density are also correlated with $M_{\rm s}$ \citep{2010ApJ...708.1204B}. Its application to Small Magellanic Cloud revealed the $M_{\rm s}$ distribution \citep{2010ApJ...708.1204B}. The aforementioned Tsallis Statistics \citep{2010ApJ...710..125E} is also sensitive to $M_{\rm s}$, so is the genus analysis proposed and applied to Small Magellanic Cloud in \cite{2008ApJ...688.1021C}. The latter measures the number of isolated islands and holes in the observed column density distribution as a function of the density threshold. Numerical simulations showed that this difference strongly depends on $M_{\rm s}$. 

The hierarchical structure induced by MHD turbulence in PPV space can be measured by Dendrograms \citep{1992ApJ...393..172H,2008ApJ...679.1338R}. In \cite{2013ApJ...770..141B}, the dendrogram analysis was introduced. Similar to \cite{2008ApJ...688.1021C}, the measures of the hierarchical structure observed in PPV were analyzed as a function of the intensity threshold to reveal their dependence both of $M_{\rm s}$ and $M_{\rm A}$. This makes the dendrogram analysis another tool to check the self-consistency of $M_{\rm s}$ and $M_{\rm A}$ values obtained with other techniques.

More recently, \cite{2016ApJ...827...26B} proposed using the Phase Coherence Function to obtain $M_{\rm s}$. This approach, similar to the bispectrum, uses the phase information of the density distribution. This makes this approach particularly sensitive to $M_{\rm s}$. 

It is worth mentioning that the velocity gradients can also provide the values of $M_{\rm s}$. This was demonstrated in \cite{2020ApJ...898...65Y}, where the dependence of velocity gradient amplitudes on $M_{\rm s}$ was revealed through numerical analysis.

\subsection{Further work: obtaining vector $B$ distribution, the synergy of approaches and Machine Learning}
This paper deals with obtaining the 3D distribution of POS magnetic field strength using the combination of the independently obtained values of $M_{\rm s}$ and $M_{\rm A}$, i.e., the approach suggested in \cite{2020arXiv200207996L} and termed there MM2. The natural next goal is to obtain the full magnetic field vector from observations. \cite{2023MNRAS.519.3736H} discussed one possible way of doing this. They noticed that the degree of dust polarization \citep{BG15} depends on $M_{\rm A}$ and the inclination angle $\gamma$ between the LOS and the mean magnetic field direction in the medium. Therefore, combining the measurements of $M_{\rm A}$ with the dust polarization degree, a full 3D magnetic field vector can be obtained, for instance in the L1688 molecular cloud \citep{2022arXiv221011023H}. 

In this paper, we discussed the approach to obtain the POS magnetic field direction and its strength in Galactic HI. Combining this work with $\gamma$, one can obtain the 3D distribution of 3D magnetic field vectors in the Galactic disk. This can have a high impact on different brunches of astrophysical research, including understanding the role that magnetic field plays in the galactic ecosystem and understanding the acceleration and propagation of cosmic rays \citep{2002PhRvL..89B1102Y,2022FrASS...9.0900B,2022FrP....10.2799L}. Naturally, such maps will significantly impact understanding the role of magnetic fields in star formation \citep{1956MNRAS.116..503M,1966MNRAS.133..265M,MK04,MO07,Crutcher12,2014SSRv..181....1L}. Especially, the 3D magnetic field is crucial for testing the theories of star formation in magnetized ISM, including the classical ambipolar diffusion theory \citep{1956MNRAS.116..503M,1966MNRAS.133..265M} and a recently suggested theory based on the concept of reconnection diffusion \citep{2012ApJ...757..154L,2014SSRv..181....1L}. In fact, our application of this afore-discussed approach for studying the L1688 molecular cloud has provided the 3D vector magnetic field in the cloud \citep{2022arXiv221011023H}. Combining such 3D magnetic field vector measurements in molecular clouds with the 3D magnetic field vector measurements in HI will allow exploring the magnetic field's direction and strength change during star formation. Such a study can also be synergetic to the Faraday rotation approach proposed in \cite{2018A&A...614A.100T} to probe magnetic fields of molecular clouds. The large-scale 3D magnetic field of the Perseus and Orion A clouds was later derived from the combination of Faraday rotation and dust polarization \citep{2022A&A...660L...7T,2022A&A...660A..97T}. 

We note that the analysis of dust polarization degree \citep{2023MNRAS.519.3736H} is not the only way of obtaining the actual 3D vector of the magnetic field. Employing the theoretical work of \cite{2016MNRAS.461.1227K}, \cite{HLX21a} proposed the SFA to obtain 3D magnetic fields using only spectral line information in PPV data cubes. This approach has been numerically tested, and its synergy with the earlier discussed approach is a promising direction for further studies. This is especially true as polarization measurements are subject to confusion in studying magnetic fields both in molecular clouds and HI in the galactic plane. In view that the SFA is also rooted in MHD turbulence's anisotropy, this property potentially can also be utilized by VGT to get 3D magnetic fields. 

Testing our results with independent approaches is another important avenue of future research. The testing of the directions of the POS magnetic field obtained with the velocity gradients was successfully performed with starlight polarization data in \citep{2019ApJ...874...25G}. There the 3D POS magnetic field maps obtained with VGT were used to predict the polarization of stars, the distances to which were measured with the GAIA survey \citep{2018A&A...616A...1G}. These predictions were then tested against the actual polarization measurements of starlight. 

This research should be extended to include the effect of magnetically aligned dust in molecular clouds. Studies of magnetic fields in nearby molecular clouds with velocity gradients have been successfully conducted \citep{HYL20,Hu19a,2022MNRAS.511..829H,2022A&A...658A..90A,2022MNRAS.510.4952L}. We expect that the correspondence of starlight polarization prediction that is available by combining the POS magnetic field maps for HI and molecular clouds will further improve the correspondence with the actual measurements of starlight polarization. 

The distribution of magnetization of nearby molecular clouds, i.e., the distribution of $M_{\rm A}$ was obtained in \cite{Lazarian18,Hu19a}. Its mean values were successfully tested against the values of $M_{\rm A}$ measured with polarization. One can obtain the POS magnetic field strength for these molecular clouds using the MM2 approach employed in this paper. This can be compared with the results obtained using more traditional ways, e.g. DCF and its modifications \citep{1951PhRv...81..890D,1953ApJ...118..113C,2022MNRAS.514.1575C,2022FrASS...9.3556L}, or a recently proposed approach of DMA \citep{2022ApJ...935...77L}. 

The predictions of $M_{\rm s}$ are based on the empirical analysis of the numerical simulations. On the contrary, the predictions of $M_{\rm A}$ for some techniques combine the analytical and numerical results. Further analytical studies are very important, as they can cover the parameter space that is very difficult to cover with brute-force numerical studies. For instance, many of the measurements of $M_{\rm A}$ are obtained for $\gamma=\pi/2$. At the same time, a recent analytical study in \cite{2022arXiv221011023H} shows that the statistics of the projected magnetic field can be significantly affected by the change of $\gamma$. Obtaining $\gamma$ from observations is feasible; thus, it is feasible to enhance the accuracy of the way of magnetic field study that we outline here.

As we discussed earlier, our present study provides the direction for further work and searches for the synergy of different techniques for determining $M_{\rm A}$, $M_{\rm s}$, and the magnetic field strength. Different techniques must use independent channels of information that allow magnetic field measurements. This calls for constructing a comprehensive technique that will obtain both Mach numbers by combining different information channels, e.g., information on magnetic anisotropies, hierarchical structure, genus, and phase information. This can be achieved by combining what we already learned about studying turbulence from observations with the Machine Learning approach \citep{2019ApJ...882L..12P}. 

\section{Discussion}
\label{sec:dis}
\subsection{Importance of the $M_{\rm A}$ and $M_{\rm s}$ distributions} 
It is important that in our paper, we first obtained the distribution of $M_{\rm A}$ and $M_{\rm s}$ and only after that applied the MM2 approach to calculate the POS magnetic field strength. Indeed, the distributions of $M_{\rm A}$ and $M_{\rm s}$ are of importance on their own. For instance, $M_{\rm A}$ taken alone is a critical parameter for describing the propagation and acceleration of cosmic rays \citep{2002PhRvL..89B1102Y,2013ApJ...779..140X,2022FrP....10.2799L,2022FrASS...9.0900B,HLX21a,2023ApJ...942...21X}. The value of $M_{\rm s}$ is important for stochastic cosmic ray acceleration \citep{2006ApJ...638..811C}, and it is definitely crucial for understanding star formation \citep{2022ApJ...927...75A}. 

As a result, the 3D distributions of $M_{\rm A}$ and $M_{\rm s}$ are important far beyond the application to magnetic field exploration. We want to stress that finding these distributions provides a very important test for our understanding of ISM physics. This physics is incorporated as subgrid models in the cosmological code describing the extragalactic evolution. Testing these subgrid models is essential for the codes to describe to be accurate. 

\subsection{Reliability of our approach}
The cornerstone of our approach in this paper is the ability to obtain the magnetic field direction and the magnetization using velocity gradients. The latter technique is based on the modern theory of MHD turbulence \citep{2019tuma.book.....B} and the theory of mapping velocity fluctuations into PPV space \citep{LP00,2016MNRAS.461.1227K,2017ApJ...838...91K}. Both theories are extensively tested numerically. 

The accuracy of the VGT in obtaining magnetic field direction and magnetization was tested with numerous synthetic observations \citep{LY18a,HLS21,2023arXiv230113458H} as well as through comparing with observed polarization data \citep{HYL20,2019ApJ...874...25G,Hu19a,2022MNRAS.511..829H,2022A&A...658A..90A,2022MNRAS.510.4952L}. VGT, as well as  intensity gradient technique, has been successfully applied to mapping magnetic field on a variety of scales, from individual molecular clouds \citep{Hu19a,HLS21,2022A&A...658A..90A,2022MNRAS.510.4952L}, Central Molecular Zone \citep{2022MNRAS.511..829H,2022MNRAS.513.3493H}, Seyfert galaxies \citep{2022ApJ...941...92H,2023MNRAS.519.1068L} , to clusters of galaxies \citep{2020ApJ...901..162H}. 
VGT has also been applied to HI \citep{YL17a,HYL20,2020MNRAS.496.2868L} with the results of VGT-obtained magnetic field maps compared to the Planck dust polarization data. 

Despite these successes, the accuracy of the VGT in studying magnetic fields in HI is sometimes debated. This can possibly be because the applicability of the VGT to HI data was challenged in \cite{2019ApJ...874..171C}. Instead, an alternative way of mapping the magnetic field based on tracing the actual cold-density filaments was advocated there. The corresponding technique, termed Rolling Hough Transform (RHT; \citealt{2014ApJ...789...82C}), attributes the elongated intensity features observed in channel maps to cold gas density filaments and, based on the Planck dust polarization measurements, suggests that these cold filaments should always be aligned with the magnetic field. Like VGT, RHT technique was applied to HI data to map Galactic magnetic fields \citep{2019ApJ...887..136C}. 

The arguments in \cite{2019ApJ...874..171C} were centered on whether the \cite{LP00} theory is applicable to the multi-phase HI gas.
To address these concerns, \cite{2023arXiv230610005H} applied a new technique termed VDA to realistic HI simulation and GALFA-HI observation. By separating contributions from cold and warm gas, this technique directly demonstrated that the VGT-related concerns in \cite{2019ApJ...874..171C} were invalid. Additionally, the RHT theoretical explanation, which is based on the alignment of the cold filament and the magnetic field, is not justified. First, this explanation disregards the effect of velocity crowding in PPV space that creates intensity filaments in velocity channel maps. \cite{2023arXiv230610005H} showed that the cold neutral medium cannot avoid velocity crowding, which is naturally raised by the Doppler-shift effect. Second, the real cold density filaments can be either parallel or perpendicular to the magnetic field. The parallel orientation of filaments is only preferable in subsonic $M_{\rm s}<1$ medium \citep{2019ApJ...878..157X,2020ApJ...905..129H,2021MNRAS.504.4354B}, making density filaments unreliable for magnetic field tracing in supersonic gas, e.g., present in the galactic disk and molecular clouds. Third, \cite{2019ApJ...874..171C} used the overlapping correlation between unsharp-masked HI structures and far-infrared emission to support the cold-filament explanation. However, \cite{2023arXiv230610005H} showed that this correlation can be purely caused by velocity crowding. In general, VGT's ability to obtain magnetic field direction is based on the anisotropy of MHD turbulence. Velocity is the direct informant of the MHD statistics, while density plays a role of an indirect messenger.

Leaving aside the theoretical justification of VGT and RHT approaches to studying the magnetic field in HI gas, we can say that both techniques were demonstrated to trace the magnetic field. If we focus on the issues of relative reliability for practical magnetic field studies, we would like to stress that, unlike RHT, VGT does not require any adjustable parameters. Within the VGT, procedures are developed that isolate/separate the contributions of density fluctuations and therefore are less affected by variations of orientation of density features. In addition, VGT is supported by MHD turbulence theory \citep{GS95,LV99} and the theory of turbulence fluctuations in PPV space \citep{LP00}. This allows for gauging the accuracy of VGT for different angles $\gamma$ between the magnetic field and the LOS \citep{2016MNRAS.461.1227K} and removing parasitic contributions of fast MHD modes, as shown in \cite{2021ApJ...911...53H}. Naturally, the empirical RHT does not possess such abilities. \cite{2020arXiv200207996L,2022arXiv220806074H} further provides suggestions for spacial filtering that increase the magnetic field tracing accuracy. 

If we consider the outputs of VGT and RHT, they are also different. VGT measured over a relatively small sub-block, e.g., typically from $15\times15$ to $60\times60$-pixels size, provides the distribution of gradients that yields $M_{\rm A}$ corresponding to the sub-block. The corresponding ability has not been demonstrated with RHT. Nevertheless, the way to find $M_{\rm A}$ with RHT is expected to be similar to the way of finding magnetization with the polarization-based DCF methods, i.e., overestimation appears if in the velocity dispersion and polarization angle dispersion \footnote{Similar estimation of angle dispersion can be also achieved with the VGT-mapped magnetic field directions and can be used for finding both mean magnetizations and mean magnetic field.} are measured at different scales (see \S~\ref{sec:method}). The comparison of the accuracy of the RHT and the vGT are rare. However, a recent comparison of these two techniques in \cite{2023MNRAS.518.4466A} showed that VGT could trace magnetic fields in HI more accurately than RHT. 

For general magnetic field studies, the VGT approach that we provided in the paper must be complementary to the magnetic field studies employing other branches of the Gradient Technique (GT). For instance, the 3D distribution of the 3D magnetic field vector can provide a way to identify the direction of the arrival of the Ultra High Energy Cosmic Rays (UHECRs), opening a new window for studying high-energy phenomena \citep{2019JCAP...05..004F}. For this purpose, one is required to know not only the magnetic field in HI but also the detailed distribution of magnetic fields in the galactic halo. The latter may be obtained using the version of GT, namely, the Synchrotron Polarization Gradients (SPGs) as demonstrated in \cite{2018ApJ...865...59L}.  

\subsection{Uncertainties}
\subsubsection{Spatial distance determined by Galactic rotational curve}
In this study, we utilize the Galactic rotational curve derived by \citet{1985ApJ...295..422C} to determine the LOS spatial position of each HI cloud. However, it is important to acknowledge that the determined position is subject to two significant uncertainties. Firstly, the Galactic rotational curve (see Eq.~\ref{eq.grc}) is obtained through fitting multiple measurements, and it may deviate in local regions, particularly for $R>8$~kpc. Secondly, we calculate the distance by considering the central coordinates and central velocity of each thick HI channel (with a channel width of approximately 10 km/s). In addition, the choice of approximately 10 km/s may not perfectly isolate every individual HI cloud.

\subsubsection{$M_{\rm A}$ and $M_{\rm s}$}
$M_{\rm A}$ distribution is derived from velocity gradient orientation. This method has a solid theoretical foundation from MHD turbulence theories. The major uncertainty here is the effect of the noise on observational data and the fitting errors. As we blanked out the gradient pixels where the HI intensity is less than three times the noise level, we expect the uncertainty contributed by noise to be insignificant. 

$M_{\rm s}$ is calculated by the column density's variance. The column density, however, is transformed from HI brightness temperature assuming a constant coefficient for the transformation. The constant coefficient may erase part of the column density's fluctuation. This raises uncertainties about the obtained variance of column density.

\subsubsection{POS magnetic field strength}
The POS magnetic field strength derived from the MM2 method involves sound speed $c_s$, HI mass density $\rho$, $M_{\rm s}$, and $M_{\rm A}$. The uncertainties in $\rho$, $M_{\rm s}$, and $M_{\rm A}$ have been discussed above. $c_s$ in our analysis is assumed to be constant, as we assume warm and unstable phases dominate the HI gas with a temperature of 10000~K. This assumption is based on the fact that the volume filling factor of gas in the cold phase is lower than 10\% \citep{2021arXiv211106845H} and the column density from the warm neutral medium can be one order of magnitude larger than that of the cold neutral medium \citep{2018A&A...619A..58K}. A constant $c_s$, however, may erase the fluctuations in the obtained $B$ (see Eq.~\ref{eq.mm2}). This uncertainty is challenging to address in observation, especially in 3D. Nevertheless, we find the mean MM2-measured magnetic field strength ranges from $0.5-2.5\mu$G, which is not far way from existing Zeeman splitting measurements \citep{Crutcher12,2022arXiv220311179P}. We, therefore, expect this uncertainty to be not significant.

\section{Summary}
\label{sec:con}
This study presents a novel way to map the POS Galactic magnetic field orientation and strength, as well as $M_{\rm A}$ and $M_{\rm s}$, in three dimensions. This is achieved by analyzing the GALFA-HI survey of neutral hydrogen and determining the gas' 3D spatial position within the Galaxy using the Galactic rotational curve. The analysis targets a low galactic latitude region and synthetically utilizes the velocity gradient technique, the column density variance approach, and the MM2 technique. Our major discoveries are:
\begin{enumerate}
    \item We presented the maps of POS magnetic field orientation in 3D. The magnetic field orientation is obtained from the velocity gradient technique;
    \item We presented the maps of POS $M_{\rm A}$ and $M_{\rm s}$ in 3D. We find the HI clouds along the LOS are globally sub-Alfv\'enic and subsonic (or transonic).
    \item We found the variation of mean $M_{\rm A}$ along the LOS approximately ranges from 0.6 to 0.9, while that of mean $M_{\rm s}$ is from 0.2 to 1.5. A drop of mean $M_{\rm A}$ and $M_{\rm s}$ appears at the distance $R\approx10$~kpc. 
    \item We presented the maps of magnetic field POS strength in 3D. We find the mean magnetic field strength varies from 2.5~$\mu$G to 0.5~$\mu$G exhibiting a decreasing trend towards the Galaxy's outskirt.
    \item We discussed other possible approaches to measure $M_{\rm A}$ and $M_{\rm s}$ distributions. We also discussed the ways to synergistically use different approaches. 
\end{enumerate}
 
\section*{Acknowledgements}
Y.H. and A.L. acknowledge the support of NASA ATP AAH7546 and ALMA SOSPADA-016. Financial support for this work was provided by NASA through award 09\_0231 issued by the Universities Space Research Association, Inc. (USRA).

\section*{Data Availability}
The data underlying this article will be shared on reasonable request to the corresponding author.


\bibliographystyle{mnras}
\bibliography{example} 

\appendix
\section{Magnetic field orientation and strength}
\label{ap:A}
\begin{figure*}
\centering
\includegraphics[width=1\linewidth]{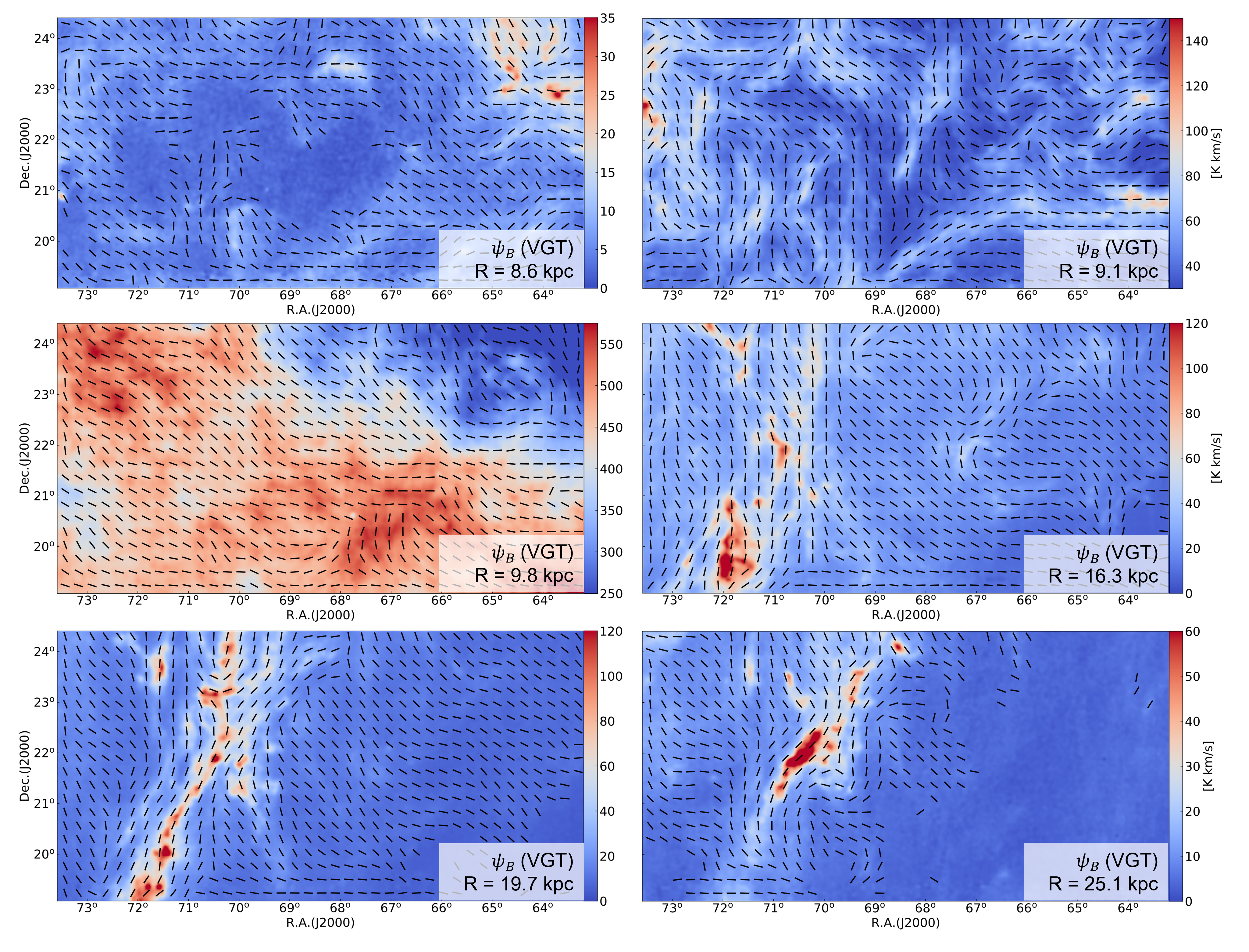}
    \caption{Maps of magnetic field orientation $\psi_B$ at six different Galactic spatial places. The magnetic field orientation (black segment) is superimposed on the integrated HI intensity maps with a channel width of $\sim10$~km/s.}
    \label{fig:Bdirection2}
\end{figure*}

\begin{figure*}
\centering
\includegraphics[width=1\linewidth]{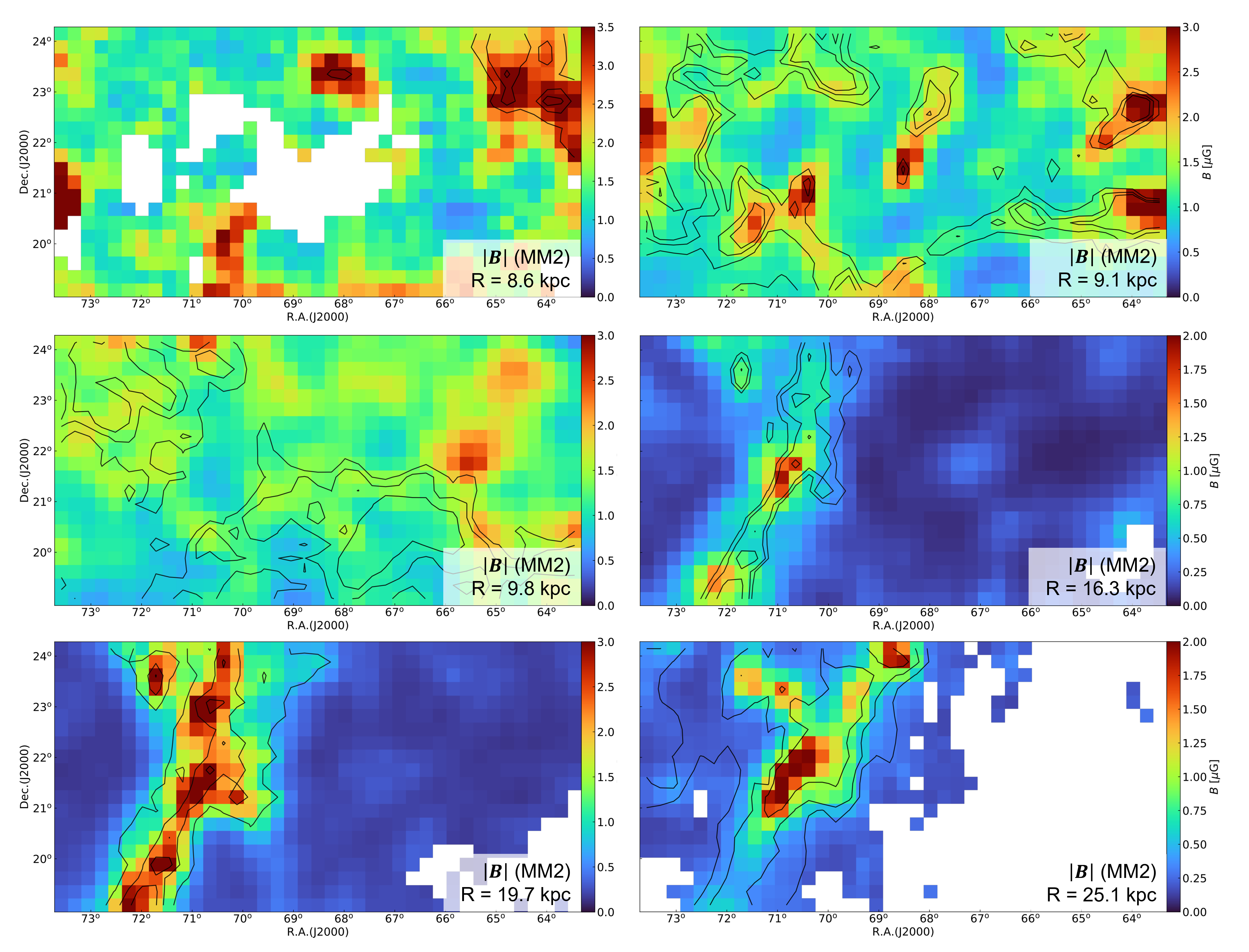}
    \caption{Same as Fig.~\ref{fig:Bdirection2}, but for magnetic field strength $|\pmb{B}|$. The contours outline the prominent HI intensity structures (see Fig.~\ref{fig:Bdirection2}) in each map.}
    \label{fig:Bstrength2}
\end{figure*}

The maps of magnetic field orientation and strength for the other six HI clouds, i.e., thick velocity channels with a width of $\sim10$~km/s, are presented in Figs.~\ref{fig:Bdirection2} and \ref{fig:Bstrength2}.

\section{$M_{\rm A}$ and $M_{\rm s}$}
\label{ap:B}
\begin{figure*}
\centering
\includegraphics[width=1\linewidth]{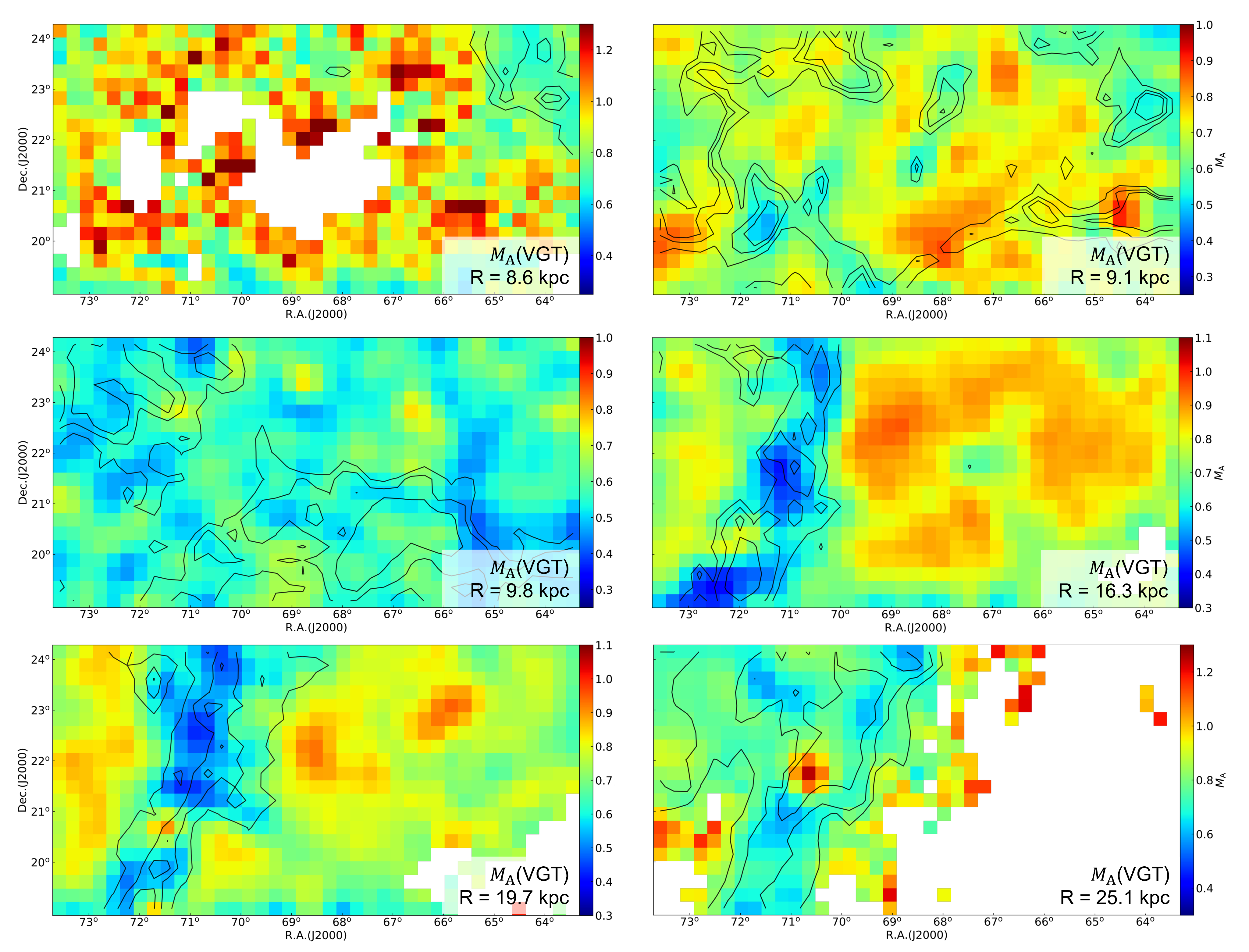}
    \caption{Same as Fig.~\ref{fig:Bdirection2}, but for $M_{\rm A}$. The contours outline the prominent HI intensity structures (see Fig.~\ref{fig:Bdirection2}) in each map.}
    \label{fig:Ma2}
\end{figure*}

\begin{figure*}
\centering
\includegraphics[width=1\linewidth]{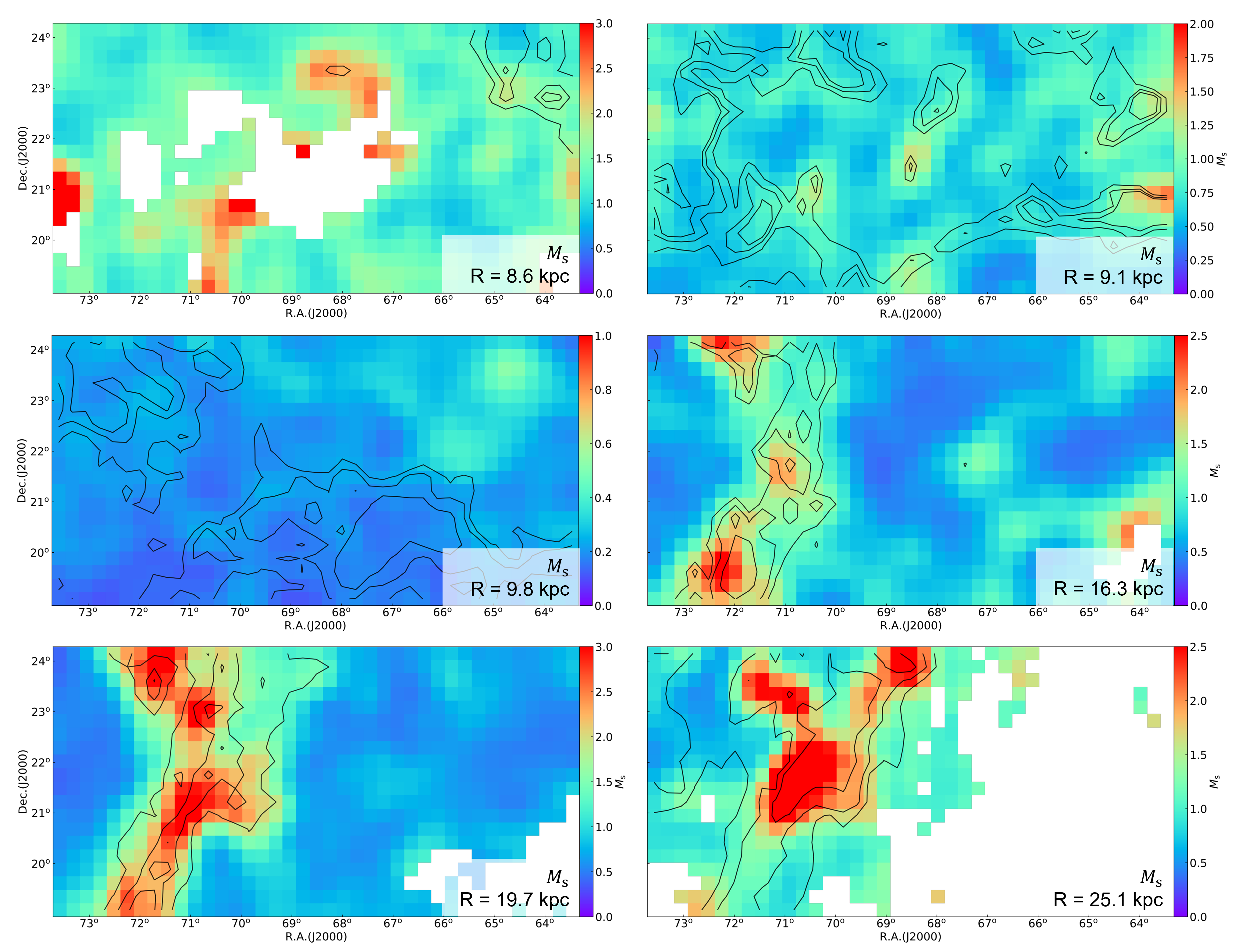}
    \caption{Same as Fig.~\ref{fig:Bdirection2}, but for $M_{\rm s}$. The contours outline the prominent HI intensity structures (see Fig.~\ref{fig:Bdirection2}) in each map}
    \label{fig:Ms2}
\end{figure*}

The maps of $M_{\rm A}$ and $M_{\rm s}$ for the other six HI cloud are presented in Figs.~\ref{fig:Ma2} and \ref{fig:Ms2}.

\bsp	
\label{lastpage}
\end{document}